# Plasma-assisted nitriding of M2 tool steel: An experimental and theoretical approach

J.M. González-Carmona[1,3], G.C. Mondragón-Rodríguez[1,3], J.E. Galván-Chaire[1,3], A.E. Gómez-Ovalle[1], L.A. Cáceres-Díaz[2], J. González-Hernández, J.M. Alvarado-Orozco[1,3]*

[1]Center of Engineering and Industrial Development, CIDESI, Surface Engineering Department, Querétaro, Av. Pie de la Cuesta 702, 76125 Santiago de Querétaro, México

[2]Centro de Tecnología Avanzada, CIATEQ A.C., Consorcio de Moldes, Troqueles y Herramentales, Eje 126 #225 Industrial San Luis, 78395, San Luis Potosí, México.

[3]Cosejo Nacional de Ciencia y Tecnología CONACyT, consorcio de Manufactura Aditiva, CONMAD, Av. Pie de la Cuesta 702, Desarrollo San Pablo, Querétaro, México.

*corresponding author(s): juan.alvarado@cidesi.edu.mx

**Abstract**

Present work was aimed to investigate the effect of nitriding time; 1, 1.5, 2.5 & 3.5 h, on surface characteristics & mechanical properties (surface nano-hardness, hardness Vickers, and adhesion) of AISI M2 tool steel. Mechanically effective compound (4 to 17.6 um) and diffusion layers (50.8 to 82.4 um) were obtained. Plasma nitriding considerably improved Surface engineering has been recognized as a family of technologies used to modify or improve the surface properties of a substrate. Nano-hardness, E-moduli and resistance to scratch of AISI M2 steel. Surface properties enhancement was associated to two harmonious effects at the nano-scale resulted by (a) stable nitride phases formation such as: $\gamma'$-$Fe_4N$, $\varepsilon$-$Fe_{2,3}N$, and (b) no brittle nitride networks and no critical decarburization of the steel upon nitriding. Thermodynamic calculations are in agreement with the experimental findings.

## 1. Introduction

Surface engineering has been recognized as a family of technologies used to modify or improve the surface properties of a substrate [1]. These have had a significant technological, economic and environmental impacts on diverse industrial sectors such as metalworking, automotive, biomedical, aeronautical, etc. In these context, thermochemical treatments such as carburizing, nitriding and nitro-carburizing are examples of large practical relevance. Thermochemical treatments are applied to tool steels providing high surface hardness, better fatigue strength, corrosion resistance and improved wear behavior along with low friction coefficient under severe working conditions of diverse mechanical systems [2–4]. Particularly, nitriding offers technical advantages over carburizing since no phase transformations are induced in the ferrite structure of steel [1]. The Iron-Nitrogen equilibrium diagram clearly indicates that N-solubility strongly depends on process temperature, and at 450°C, about 5.7 to 6.1 % nitrogen will be expected to dissolve into the iron lattice [5]. At higher temperatures (> 500 °C) nitrogen solubility decreases affecting the bulk mechanical properties of tool steels.

On the other hand, surface characteristics of the treated steel will be strongly affected by nitrogen diffusion and its control would be crucial to guarantee success during high performance operations.

Plasma nitriding process (PNP) is carried out under vacuum (approx. $10^{-2}$ mbar), moderate temperature (< 400 - 450 °C) and pulsed DC current. Potential is applied between the sample/substrate and the reactor chamber [6]. Compared to conventional nitriding processes, pulsed plasma nitriding offers a number of advantages including, low operating temperatures which cause fewer piece distortions and no changes on bulk mechanical properties. Surface roughness is not considerably affected by plasma nitriding and no finishing post-treatments are needed. In spite of this surface properties enhancements, plasma nitriding may cause nitrogen networks formation which may cause catastrophic failures (i.e. cracking, spallation, chipping) during operation [1]. Plasma nitriding may also induce critical decarburization of steels. Sun et al [2] reported the effect of substrate carbon content on the formation of nitride structures of low alloy steel and stated that process time increases decarburization but promotes $\gamma'$-$Fe_4N$ formation. The $\gamma'$-$Fe_4N$ phase is highly desirable due to its mechanical resistance. These surface modifications play an important role in applications where the nitride layer acts as a high tech support i.e. duplex layer for hard coatings. For tooling application, relatively short nitriding times are of practical interest. Nevertheless, most reports on plasma nitriding are dedicated to investigate the treatment effect on AISI 316L characteristics, i.e. 5 h at 450 - 480 °C [2,7,8], 8 h at 400°C [9], 20 h at 400 °C [10]. Long nitriding times lead to considerable higher operating costs and steel decarburization. AISI M2 tool steel is widely used in the tool industry, however, plasma nitriding results are scarcely found in the literature. For instance C. Kwietniewski et al [11] investigated the edge embrittlement of AISI M2 machining tools caused upon plasma nitriding and concluded that excessive deep and brittle diffusion zone can be induced and can be considerably higher at the tool edges. These effects are particularly enhanced when using brighter plasma nitriding. A. Akbari et al [12] investigated the effect of the initial microstructure on the thickness and hardness of the nitrided layer of AISI M2 produced by plasma nitriding at 450 °C and 8 h. They observed that previous thermal treatments have an impact on the steel microstructure and on the surface hardness plasma after nitriding.

Considering that nitriding process parameters such as bias voltage, temperature and time, may induce decarburization and promote unfavorable effects on the surface microstructure-properties, it is required to tune the process parameters and to investigate their effect on phase(s) stability, surface mechanical properties and nitrided system characteristics. In this work the effect of time during plasma nitriding on the surface properties of high alloyed AISI M2 steel was investigated, particular focus is paid on nano-mechanical properties and scratch resistance. These surface properties are correlated with the nitride phases identified via grazing incidence XRD measurements, Rietveld refinement and thermodynamic calculations applying Thermocalc.

## 2. Experimental Methodology
*2.1 The plasma nitriding process (PNP)*

Plasma nitriding process was carried out on commercial AISI M2 tool steel grade substrates, 25.4 mm diameter and 5 mm height cylindrical samples. Chemical composition of the steel was determined using a Horiba GD profiler 2, Glow Discharge Optical Emission Spectroscopy (GDOES) and the results are shown in Table I. Prior PNP surface samples were treated using abrasive paper (180 - 1200) and mirror polished applying diamond paste (3 - 6 um), then ultrasonically cleaned in isopropyl alcohol, dried and fixed in the substrate holder to be placed in the reactor chamber.

Table I. Chemical composition of commercial AISI M2 steel.

| Element | C | Mo | V | Cr | W | Co | Si | Mn | Fe |
|---|---|---|---|---|---|---|---|---|---|
| wt. % | 0.9 | 3,83 | 1,43 | 3,97 | 4,17 | 0,445 | 0,39 | 0,286 | Balance |

For Plasma nitriding, a Metaplas Domino Mini system from Oerlikon was used. The unit is equipped with an Arc Enhanced Glow Discharge (AEGD) process in a shielded titanium target to generate high electron and ionic density in the chamber [13], using an 80 % active RF cathodic arc with 80 A at 20 kHz. Plasma cleaning during 25 min was performed in Ar atmosphere with a -250 V RF bias at 20 kHz and 80 % active time, to reduce superficial contamination and activate the substrate surface. The samples were heated to about 350 °C prior to nitriding using a 9 kW radiator. Then, the samples were plasma nitrided during 1, 1.5, 2.5 and 3.5 hours. Table II shows detailed information regarding the nitriding parameters used in this investigation.

Table II. Plasma nitriding parameters for AISI M2 tool steel samples.

| Background pressure (mPa) | Working pressure (Pa) | Temperature (°C) | Ar:N2:H2 flow (sccm) | Polarization voltage (20 kHz/80 % active) (V) | Process times (h) |
|---|---|---|---|---|---|
| 4 | 1 | 350 | 278:50:25 | -250 | 1, 1.5, 2.5 & 3.5 |

*2.2 Microstructural and phase analysis after nitriding*
Microstructural analysis was performed using an Optical Microscope (OM). For this purpose, the samples were cut using a diamond disc and mounted in Bakelite. The fixed cross-sections of the specimens were mirror polished and etched using a 4 vol. % Nital solution. Crystalline structure changes of the samples after nitriding were followed by mean of Grazing Angle X-Ray Diffraction (GAXRD) in a Rigaku Smartlab diffractometer, using a Cu-K∝ (λ= 1.54059 Å) radiation source, in a 2θ range between 10º and 90º with 0.01° step/s. Phase quantification was performed by Rietveld Refinement analysis considering the Fundamental Parameters Approach (FPA) of the indexed nitrides and carbides applying the PDXL software from Rigaku, obtaining in all cases correlation values in the least squares method of Rwp below 3 % and normalized residual (S) < 1.09.

*2.3 Computational Thermodynamics approach*
Experimental changes of the parent phase and precipitates of AISI M2 steel as a function of nitriding time were investigated applying the CALPHAD methodology [14] and using the

2016b version of the ThermoCalc software. The initial state of equilibrium of the M2 steel was studied based on standard values of the state variables and the effects of N on nitrides formation and evolution of carbides were followed by single point equilibrium and mapping calculations with the TCFE8 Iron base alloys database [15]. The experimental conditions; chemical composition of the steel (see table I) and nitriding process parameters (see table II) and system size were considered as input data for the calculations.

*2.4 The nano-mechanical and scratch testing evaluation*

Nano-mechanical properties of the plasma nitrided surfaces were measured by instrumented nanoindentation in a Hysitron TI-700 UBI equipped with a Berkovich indenter, the Oliver and Pharr method was used to calculate hardness (H) and elastic moduli (E) by applying a 5000 µN constant load. Corrections were performed in compliance, contact area and thermal drift using fused silica as a reference. Weibull distribution analysis was assessed to assure representative evaluation of the surface properties [16]. Vickers microhardness measurements in the cross sections were performed using a Vickers diamond indenter applying 1 N load.

Failure mechanisms and scratch resistance of the nitrided layers were evaluated in an Anton Paar Revetest Scratch tester. Scratch resistance was evaluated by a dynamic scratch method and three tests average, see test conditions in Table III. Prior scratch, the instrument performs a pre-scan on the surface applying 0.01 N load. Then load gradient is applied to start at 0.1 N up to 35 N. After that, a post-scan is applied on the worn track using the same load. Therefore, the Residual depth (Rd) is calculated as the difference between the scans and maximum penetration depth are selected to analyze scratch resistance. Track characteristics and the failure mechanisms were analyzed by optical microscopy.

**Table III. Scratch test parameters of plasma nitrided M2 steel surfaces.**

| Indenter | Rockwell C |
|---|---|
| Load gradient (N) | 0.1 – 35 |
| Scratch speed (N/min) | 14 |
| Distance (mm) | 3.5 |
| Speed (mm/min) | 11 |

**3. Results and discussion**

*3.1 Compound and diffusion layers*

The effect of the plasma assisted nitriding time on the compound (CL) and diffusion layers (DL) of the M2 steel is shown in the optical images of Fig. 1. In all specimens upon increasing treatment times (1.0, 1.5, 2.5 & 3.5 h) two regions are observed which clearly display N gradients starting at the surface and going up to the alloy core. The first region (compound layer), consist in a thin layer of $\gamma'$-$Fe_4N$ and $\varepsilon$-$Fe_{2,3}N$ phases with fine precipitates of nitrides and carbonitrides of alloying elements [12,17]. The second region defines the diffusion layer which is subdivided into two zones; the first zone under the compound layer consisting mainly of $\gamma'$-$Fe_4N$ phase and $M_6C$ carbides (where M = W, Mo, Co) precipitated in the grain boundaries of α-Fe matrix & the second zone having alloying precipitates with continuous reduction of the γ′-$Fe_4N$ phase [12,17]. This last nitrided "third" layer corresponds to the transition zone between the nitrided layer and the steel core. Compound layer formation on high alloy steels depends on nitrogen dissociation and ionization rate, which increases when nitrogen partial pressure and ion bombardment energy drop. Compound layer growth is also affected by sputtering during plasma nitriding, which proportionally increases with the argon partial pressures inside the reactor chamber [2,4]. The compound and diffusion layers obtained in the present investigation indicate that nitrogen potential, which is time and temperature dependent, overcomes the threshold needed to obtain high nitrogen ionization rate and high energy for ion bombardment resulting in a well-defined compound layer and the subsequent diffusion layer. Thickness layers are average values of 5 measurements.

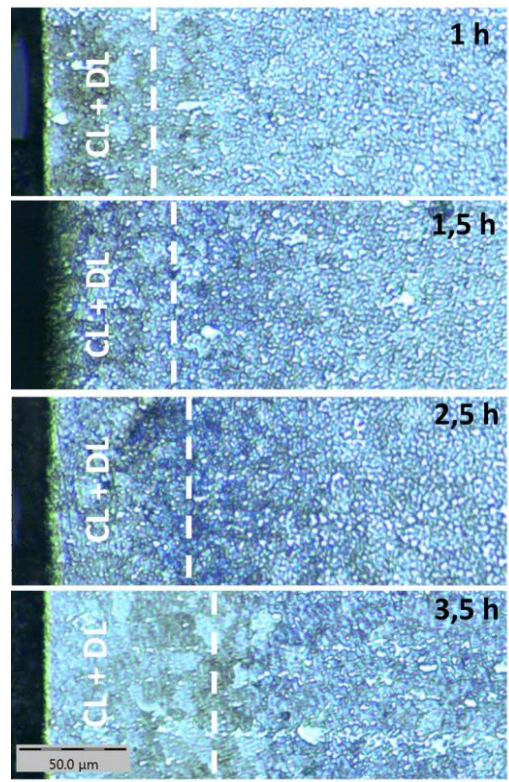

Figure 1. Optical microscopy of cross-sections profiles of plasma nitriding M2 steel at different nitriding times and 350 °C.

The effect of nitriding time on compound- & diffusion-thickness layers are presented in Fig. 2. This graph shows an approx. 4 μm thick compound layer (CL) reached only within 1 h. Nitriding time increment to 1.5 h resulted in about 66.1 % CL-thickness increase. Only ~ 69.2 % increase on CL-thickness was achieved upon 2.5 h nitriding while about 77.4 % relative increase was obtained after 3.5 h (17.6 μm). Similar effect was observed on the diffusion layer (DL) thickness where ~ 50.8 μm was reached within 1 h. Increments of 15.3 % and 39.4 % DL-thicknesses were observed after 1.5 and 2.5 h nitriding times, respectively. A maximum DL-thickness increment of about 62 % (82.4 μm) was reached after 3.5 h. These CL and DL thickness results highlight the efficiency of the plasma nitriding process. Less than 10 um thick compound layer was reached by Avelar-Batista et al [4] after 4 h plasma nitriding of AISI M2 in 40 % Ar + 60 % $N_2$ (-500 V bias and 500 °C). The use of higher nitrogen concentration (70

%) led to almost no compound layer formation, but having a slight detrimental impact on Vickers micro-hardness. In our investigation the use of hydrogen and the AEGD process enhances surface activation and ion density in the reaction chamber, reducing even further the time needed to obtain well defined compound and diffusion layers in high alloyed AISI M2 steel.

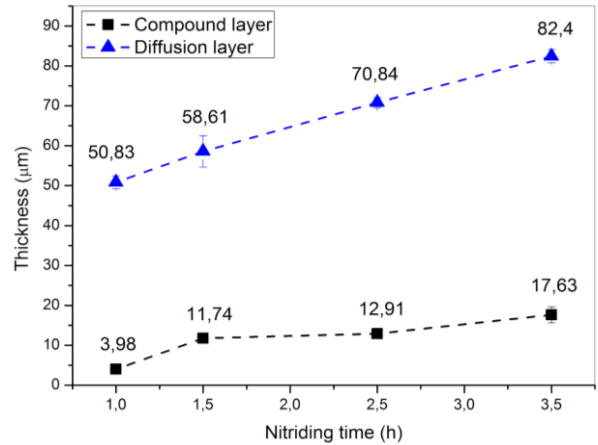

Figure 2. Compound and diffusion layer thicknesses after plasma nitriding at different process times.

*3.2 The crystalline structure and phase stability*

The changes in the crystal structure of the commercial AISI M2 steel surface before and after plasma nitriding process are shown in the grazing incidence XRD patterns in Fig 3a. The crystalline structure of the Non-Nitrided sample is composed of the α-Fe matrix and dispersed complex carbides identified as $M_6C$ and MC. Carbides formation in AISI M2 steel has been reported elsewhere [12,18] and is expected due to the stabilizing effect of C combined with alloying transition metals such as Cr, Mo, W, etc, which are reported in the Table I. The occupation of the d level in the electronic structure of alloying elements plays a determinant role. The less occupied the d level, the greater the probability of forming carbides. General, the most common carbides reported for the M2 steel are identified by XRD in this work. Other rare carbides such as $M_{23}C_6$ and $M_7C_3$ and $M_3C$ were not observed. As the nitriding time increased the XRD patterns allowed to identify the presence of the ε-$Fe_{2,3}N$ and γ′-$Fe_4N$ iron nitrides, which are known as hard-crystalline phases that form the compound layer (also known as white layer) upon the plasma assisted nitriding process. The changes in the diffraction peaks of the α-Fe matrix and those combined for Carbides and Nitrides allow to observe an evolution of the M2 steel crystal structure as a function of nitriding time and are quantified by means of Rietveld refinement.

A detailed inspection of the range from 43° to 46° is presented in the Fig 3b, the integral peak shifting to lower angles can be associated to the contribution of increasing in lattice constant of the α-Fe phase and uniform macro stresses in the compound layer, while the peak broadening is related to non-uniform stresses and the combined contribution of α-Fe and γ′-$Fe_4N$ phases. This increase in lattice constant is caused when nitrogen diffuses into the crystalline structure of the α-Fe and it locates in the interstitial sites. Additionally, macro-stress are associated to anisotropic nitrogen diffusion which causes nitrogen gradients inside grains and subsequent thermal expansion mismatch of the nitrided layers and steel bulk. These two factors produce strain fields in the matrix around the nitrides precipitates which are the main causes for micro and macro-stress generation upon surface modification.

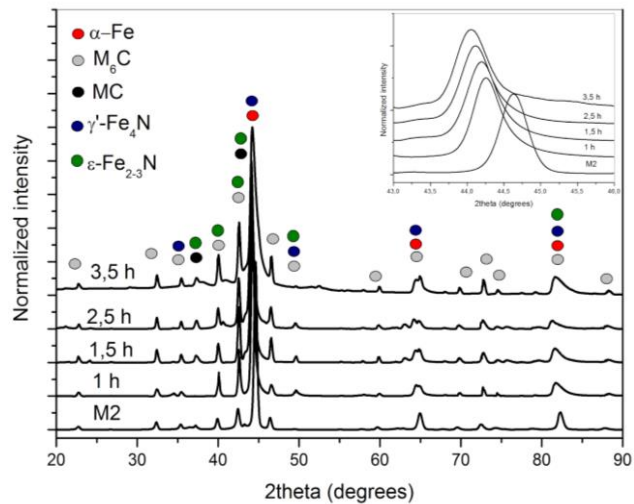

Figure 3. Grazing incidence XRD patterns of the AISI M2 steel before and after plasma nitriding as a function of nitring time, and α-Fe peak shift after nitriding (see insert).

*3.2 Rietveld analysis and phase calculations*

The phase content of α-Fe matrix, carbides and nitrides obtained by Rietveld refinement as a function of nitriding time is shown in Fig. 4. It was found that the Non-nitrided M2 steel is composed of 77 wt. % of α-Fe, 19 wt.% of $M_6C$ and 4 wt.% of MC carbides. As the nitriding time increases the α-Fe phase content decreased linearly to 30 wt.% at 3.5 h while the $M_6C$ exhibited initially a decreased to a minimum of 3.5 wt.% at 1 h and later slightly increased to 4.5 wt.%, 6.5 wt.% and 7.5 wt.% at 1.5, 2.5 and 3.5 h, respectively. In turn, the MC content exhibited an increase from 4% to 21 %. The appearance of the $\gamma'$-$Fe_4N$ and $\varepsilon$-$Fe_{2,3}N$ nitrides show the success of the nitriding process with increment of the content from 0 to 30 wt.% of $F_{2,3}N$ at 3.5 h. An interesting behavior of the $\gamma'$-$Fe_4N$ content was observed, increasing to 27 wt.% at 1.5 h and 2.5 h and then decreasing to 10 wt.% at 3.5 h. In general, the ratio of α-Fe to carbides phase content (77/23) for the non-nitrided M2 steel was found to be around 3.35. These results allow to explore the evolution of the matrix phase of the steel and the phase transformations into $\varepsilon$-$Fe_{2,3}N$ and $\gamma'$-$Fe_4N$ nitrides and probably carbonitrides. This phenomena have been observed when nitriding high alloyed ferritic steels and is related to the carbon diffusion toward the surface during the nitriding process, in which the nitrogen atoms replace the carbon atoms in the interstitial sites of the crystalline structure of the carbides to form nitrides, leaving them free to diffuse and even to sputter out of the material when the nitrogen activities are high. The increasing in lattice constant observed in Fig. 3b and the decrease in the α-Fe content in the compound layer observed in Fig. 4 is the result of N inclusion in the interstitials sites in the ferrite and heterogeneous N enrichment that leads to local equilibrium and subsequent transformation to $\gamma'$-$Fe_4N$ and $\varepsilon$-$Fe_{2,3}N$. This is consistent with the compound layer microstructure reported in the literature [12,17]. The decrease in the $M_6C$ content is associated to partial decarburization in the early stage of the nitriding process as the nitride phases stabilize [2], which suggests along with the maximum values of the $\gamma'$-$Fe_4N$ phase that there are critical conditions for equilibrium of the matrix, nitrides and carbides in the compound layer. This behaviour is described following based on thermodynamic calculations.

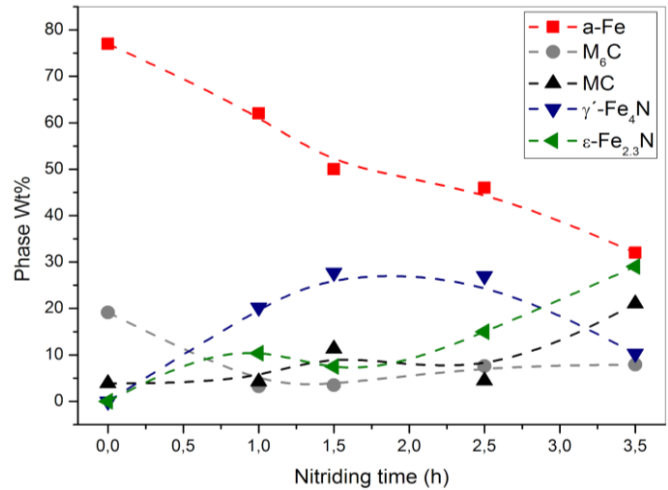

Figure 4. Effect of time on crystal phase content calculated after Rietveld Refinement of plasma nitrided AISI M2 steel.

*3.3 Thermodynamic study, the effect of N on phase precipitation in M2 steel.*
The equilibrium phases in the AISI M2 steel as a function of N content is shown in Fig. 5. The phase fraction obtained for the non-nitrided sample is composed of the α-Fe steel matrix with 80 wt.% and complex carbides carbides $M_6C$, $M_{23}C_6$ and MC with values of 11, 6.7 and 2.1 wt.% respectively, as indicated in the zoom insert in Fig. 5. The calculated ratio α-Fe/carbides (80/20) for the non-nitrided M2 steel is 4 and it is comparable to the experimental data obtained by Rietveld refinement shown in Fig. 4. A strong dependence of the thermodynamic stability of the former compound layer phases with the N content was observed. In the range of N content ($X_N$) $X_N < 2$ wt. %, the fraction of α-Fe remains approximately constant and the precipitates $M_6C$, MC and $M_{23}C_6$ begin to decrease upon N increase. In the same way it is identified as the metallic nitrides MN start to precipitate at the expense of the carbides. This decrease of the content of stable carbides as a function of N explains the decarburization phenomenon observed in the first stage of the nitriding process (up to 1 h) in the XRD patterns. In the composition range of $2 < X_N < 4$ wt. % the metallic nitrides MN and carbonitrides M(C,N) stabilize while the α-Fe phase content decrease to 70 wt.%. The observation of the sublattice model of α-Fe described in the calculations and presented in table IV, allow to identify that this phase does not accept N in interstitial sites and instead it considers vacancies Va, then this phase is labeled as α-(Fe,Va).

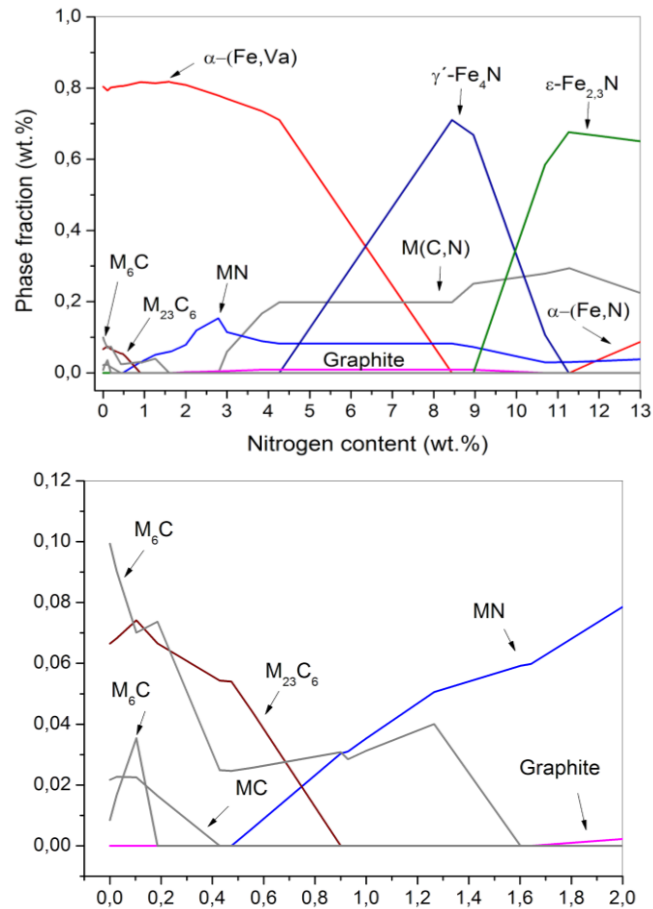

Figure 5. Calculated equilibrium diagram vs nitrogen concentration for AISI M2 steel at 350 °C.

It is observed that the γ′-Fe$_4$N phase appears in the range of 4 < $X_N$ < 9 wt. %. According to the equilibrium calculations, this phase increases its content up to a maximum value of 70 wt. % for a N critical value of 8 wt. %. It can be clearly observed that the N content stabilizes the cubic phase γ′-Fe$_4$N and accelerates the phase destabilization of the α-(Fe,Va). As the N content increases, it can be seen that metal nitrides MN and carbonitrides M(C,N) are stabilized and their phase content remains approximately constant. So, it can be stated that in this composition range the equilibrium balance occurs between the α-(Fe,Va) and γ′-Fe$_4$N phases.

For 9 < $X_N$ < 11 wt. % we observe that the ε-Fe$_{2,3}$N phase stabilizes. It can be said that this phase with hexagonal structure grows uniquely at the expense of the cubic γ′-Fe$_4$N, since the other present phases remain constant in that composition range. It was found that for For $X_N$ > 11 wt. %   For 9 < $X_N$ < 11 wt.%   the a-Fe phase stabilizes with N inclusions interstitially, which is represented by the model α-(Fe,Va). This inclusion suggests an increase in the lattice constant. According to the calculations, the N-increase favors the stability of the phases α-(Fe,Va) and γ′-Fe$_4$N before stabilizing the phase α-(Fe,N). This is in discrepancy with that observed in the XRD patterns for 1 and 1.5 h of nitriding, which indicate an increase in the lattice constant of the matrix phase, so it is suggested that the experimentally stable phase is α-(Fe,N) in the presence of the γ′-Fe$_4$N phase. These results of equilibrium calculations allow the identification of N composition regions in which ideal phase transformations occur and one phase clearly increases at the expense of another that stabilizes. But when compared with experimental results, it can be identified that the nitriding conditions promoted localized equilibrium of γ′-$Fe_4N$, ε-$Fe_{2,3}N$, α-Fe with interstitial N, metal nitrides and carbonitrides. This suggests that in the conditions of the nitriding process it is not necessary to have a total equilibrium, as is the case of thermodynamic calculation with ThermoCalc. During the experiment it can happen that we have zones of heterogeneous concentration of N and subsequently partial or local equilibrium zones where the phases observed by XRD predominate. The discrepancy between the quantification of experimental phases by XRD and the calculation of thermodynamic equilibrium by ThermoCalc can be explained considering that in the nitriding process there are surface activation energy conditions, partial pressures of process gases and heterogeneous diffusion of N in the matrix α-Fe reinforced by carbides and emerging nitrides, which increases the complexity of the system and cannot be reproduced under the equilibrium conditions of the program. This particular phenomenon should be studied in future works where ideal experiments with model alloys are defined and study the long-range N diffusion profiles in α-Fe matrix including models for dispersed phases such as the Labyrinth model. These results allow to associate the conditions of the nitriding parameters on the stability of the phases in M2 steel. It can be identified that there are stages in the nitriding process in which there is decarburization but then begins to form carbonitrides that are stable in combination with the phases γ′-$Fe_4N$ and ε-$Fe_{2,3}N$, which is convenient for the thermodynamic stability of the white layer and the strengthening of the mechanical properties of the nitrided component.

**Table IV. Crystal structures with their respective sublattice model and its chemical formula.**

| Crystal Structure | Sublattice model | Formula |
|---|---|---|
| BCC | $(Cr,Fe,Mo,V,W)_1 (C,N,VA)_3$ | Fe, Va |
| C | $(C)_1$ | C |
| FCC_A1#2 | $(Cr,Fe,Mo,V,W)_1 (C,N,VA)_1$ | MN |
| HCP_A3 | $(Cr,Fe,Mo,V,W)_1 (C,N,VA)_{0.5}$ | $MN_{0.5}$ |
| FE4N_LP1 | $(Cr,Fe)_4(C,N)_1$ | $Fe_4N$ |
| FCC_A1 | $(Cr,Fe,Mo,V,W)_1 (C,N,VA)_1$ | CrN |
| MC_SHP | $(Mo,W)_1 (C,N)_1$ | (MO,W)C |
| $M_{23}C$ | $(Cr,Fe,V)_{20} (Cr,Fe,Mo,V,W)_3 (C)_6$ | $M_{23}C$ |
| $M_6C$ | $(Fe)_2 (Mo,W)_2 (Cr,Fe,Mo,V,W)_2 (C)_1$ | $M_6C$ |

*3.4 Mechanical properties*
Nano-indentation experiments were performed to measure mechanical properties on the nitrided surfaces. Surface nano-hardness and elastic modulus vs. nitriding time are reported in Fig. 5. Nano-hardness increases up to 255 % with the maximum 3.5 h nitriding time. This effect is mainly due to formation of stable nitrides (mainly γ′-$Fe_4N$ and ε-$Fe_{2,3}N$ ) and the local microstructural changes in the nitrided zone caused upon nitriding. Large nitride grains surrounded by narrow grain boundaries are produced, and this microstructural arrangement avoids dislocations movement, preventing plastic flow. The elastic modulus and increased hardness indicates better resistance to plastic deformation of the nitrided surface. Nitride precipitates in the α-Fe matrix and the uniform stress gradient observed by XRD resulted in phase dispersion hardening and explained the good mechanical response of the nitrided system.

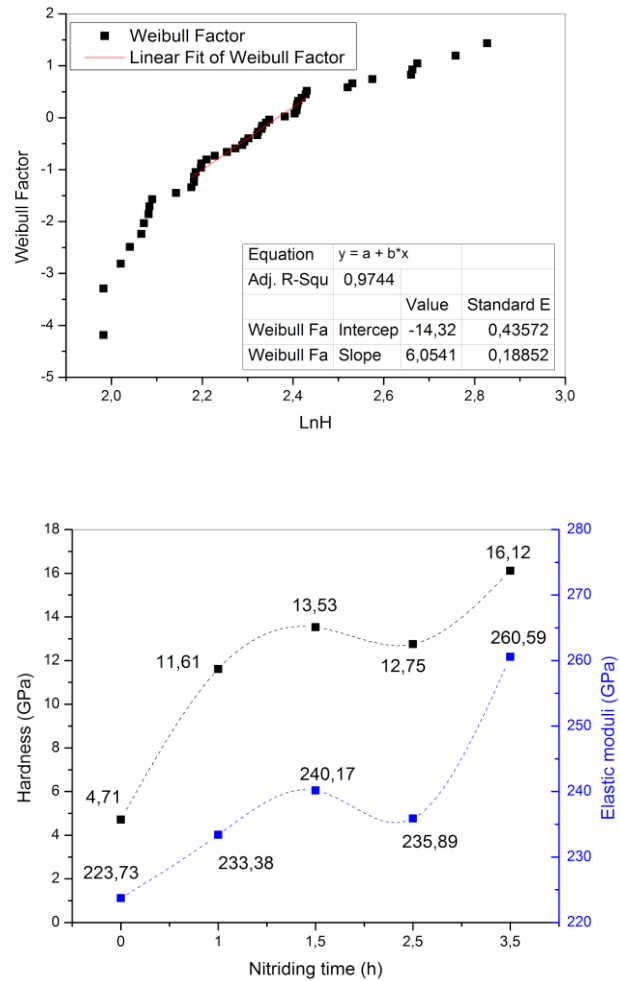

Figure 6. Effect of plasma nitriding time on surface nano-hardness & Elastic modulus of AISI M2 steel.

Hardness Vickers has be used as criteria to determine the diffusion zone thickness and the boundary to the steel core [3,12]. In the present study, microhardness measurements performed on the cross-sections at increasing surface depths are shown in Fig. 7. Nitrided surfaces showed similar hardness (600 - 900 HV) up to 40 μm depth, regardless the nitriding times. This corresponds to a 460 % increase relative to the steel without treatment. Upon 3.5 h nitriding time, high hardness is obtained up to depths close to 80 μm, which is in good agreement with the findings made by optical microscopy. At depths > 120 μm hardness is related to the non-treated steel (192.3 ± 5 HV). This hardness increase is related to formation of stable nitrides ε-$Fe_{2,3}N$ and γ′-$Fe_4N$ identified by the GAXRD analysis. These nitride phases, the solid solution of N atoms in the α-Fe matrix and alloy nitrides precipitates, cause grain size increases and reduces grain boundaries, acting as dislocation movements blockers and residual stresses produced at the nitride layers [12][19].

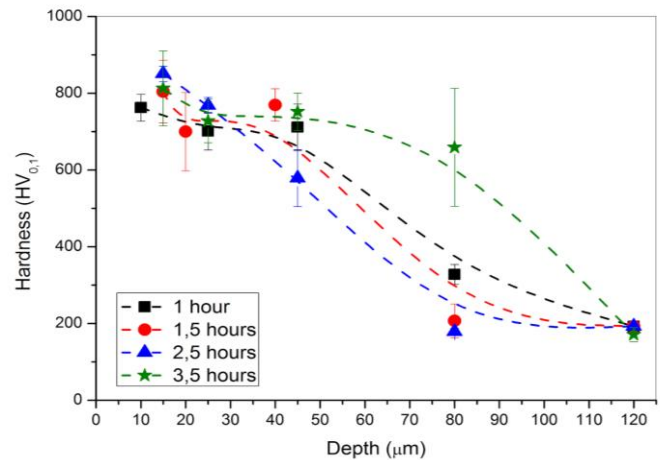

Figure 7. Hardness Vickers on the cross-sections of nitrided M2 steel coupons.

*3.5 Scratch resistance*

The scratch results were highly reproducible and reported data were averaged values of three measurements. Non-treated M2 steel surface shows mainly plastic deformation, observed during surface plowing as the indenter progresses. The worn surface shows no signs of cratering, debris or spallation. Regarding the nitrided steels, circumferential cracks are observed on coupons nitrided during 1, 1.5 and 2.5 h. These failures appear at even shorter distances upon nitriding time increase: 2,1 mm (21.2 N), 1.8 mm (18.6 N) and 1.6 mm (16.5 N) respectively. This behavior is in concordance with the nano-mechanical properties of the surface. Plasma nitriding improves nano-hardness and elastic modulus causing an increment of resistance against plastic deformation. This is related to the fracture toughness decrease due to iron nitride layers formed upon nitriding.

Details about the onset cracking and the last stage of the scratch tests on the nitrided surfaces are shown in Fig. 8. Surfaces nitrided during 1, 1.5 and 2.5 h display tensile cracks indicating that failure modes are mainly defined by tensile stresses occurring in the back of the indenter. This also suggests that cracks come from relatively brittle surfaces (iron nitrides) [20] [21]. Similarly, as shown in Fig. 9 the maximum depth of the groove only exceeds thickness of the composite layer build after 1 h of nitriding time, which indicates that formed nitrides at these treatment times are fragile and have relatively low toughness.

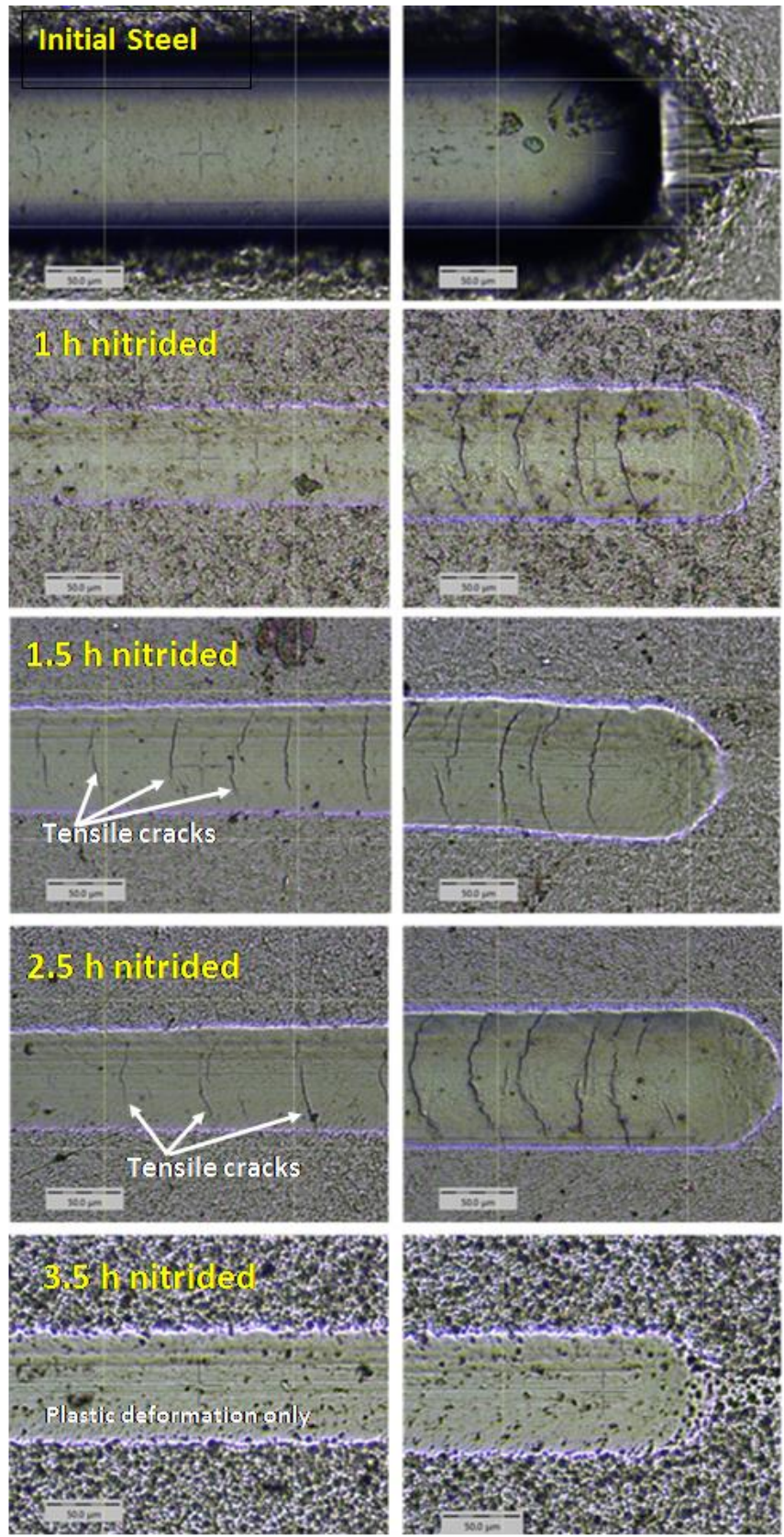

Figure 8. Onset cracking after scratch testing of initial AISI M2 and after plasma nitriding.

Cracking on the nitrided surface was not observed in the sample treated during 3.5 h. Additionally, the lowest penetration depth in the scratch pattern was reached on this surface.

Highly homogeneous microstructure (see profile in the cross-section image, Fig. 1), achieved after 3.5 h nitriding treatment, explained this great mechanical response. This system has the highest capacity to withstand scratch load and plastic deformation which is the predominant failure mode despite displaying the highest surface nano-hardness. At this point a positive balance between nitrided phases and mechanical properties was reached. Hard nitride phases ($\varepsilon$-Fe$_{2,3}$N), carbides, probably carbonitrides and adequate matrix gradients (i.e. crystalline phases and hardness) from the surface to the core of the ferrite and carbide precipitates in the steel matrix are in nano-mechanical harmony. Although the $\varepsilon$-Fe$_{2,3}$N phase upon nitriding time develops, under the selected process conditions formation of fragile iron nitride-networks on the steel surface upon 3.5 h plasma nitriding is avoided.

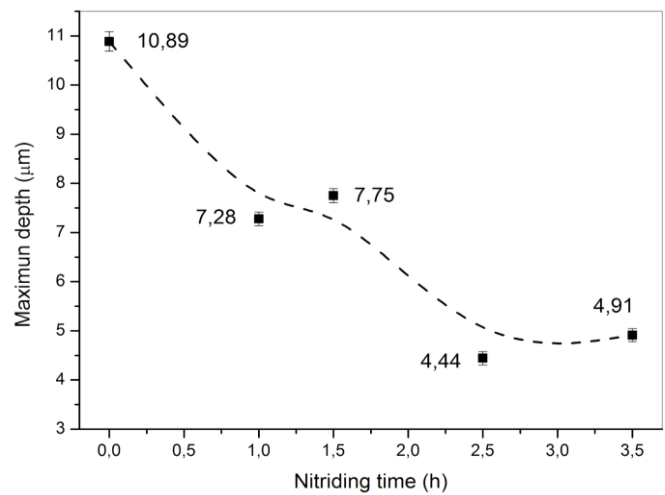

Figure 9. Maximum penetration depth after scratch-testing on initial AISI M2 and after plasma nitriding.

## 4. Conclusions

Effect of plasma nitriding time of M2 steel surface characteristics was investigated. Enhancement on surface properties upon increasing process time during plasma nitriding has been proven. Upon nitriding nano-hardness & E-moduli improves considerably and this positive effect was already visible after 1 h process time. Same trend was observed for 1.5 and 2.5 h nitriding treatments. Increments of about 255 % and 14.1 % on nano-hardness and E-moduli were reached at the maximum 3.5 h nitriding time. Scratch resistance of nitrided surfaces improves upon increasing treatment times as well. Minor embrittlement of the surface was observed upon nitriding up to 2.5 h and, surprisingly, almost negligible effect on the nitrided surface after 3.5 h. Surface resistance improvements are associated to precipitation and formation of well mechanically and chemically bonded compound and diffusion layers containing stable iron nitrides $\gamma'$-Fe$_4$N + $\varepsilon$-Fe$_{2,3}$N, and carbides in the ferrite matrix. XRD phase calculations indicate slight steel decarburization upon nitriding, but, the nano-mechanical response of the nitrided steels indicates that no critical carbon depletion. Plasma nitriding conditions suggest a highly suitable interface-support for duplex applications at process times up to 3.5 h.

Acknowledgments

The authors thank financial support provided by CONACyT (Consejo Nacional de Ciencia y Tecnología) through the Program "Frontiers of Science" under the project 1077-02-2015. Nano-indentation experiments were carried out at the Institute of Materials Research and the Ceramics department at the Universidad Michoacana de San Nicolás de Hidalgo. Thanks to Dr. Juan Zárate Medina & M.Sc. Héctor Orozco for technical support.

Thanks to Ing. VM. Hurtado Pájaro for his contribution, helping during sample preparation and his effort during nitriding.